\documentclass[twocolumn]{aastex631}

\usepackage{longtable}
\usepackage{amsmath}
\usepackage{xcolor}
\usepackage{natbib}
\usepackage{CJK}
\usepackage{nccmath}

\shorttitle{Count models for GC populations}
\shortauthors{Berek et al.}

\begin{document}

\title{Should zeros count? Modeling the galaxy-globular cluster scaling relation with(out) zero-inflated count models}

\correspondingauthor{Samantha Berek}
\email{sam.berek@mail.utoronto.ca}

\author[0000-0001-7549-5560]{Samantha C. Berek}
\altaffiliation{Data Sciences Institute Doctoral Student Fellow}
\affil{David A. Dunlap Department of Astronomy \& Astrophysics, University of Toronto, 50 St George Street, Toronto, ON M5S 3H4, Canada}
\affiliation{Dunlap Institute for Astronomy \& Astrophysics, University of Toronto, 50 St George Street, Toronto, ON M5S 3H4, Canada}
\affiliation{Data Sciences Institute, University of Toronto, 17th Floor, Ontario Power Building, 700 University Ave, Toronto, ON M5G 1Z5, Canada}

\author[0000-0003-3734-8177]{Gwendolyn M. Eadie}
\affiliation{David A. Dunlap Department of Astronomy \& Astrophysics, University of Toronto, 50 St George Street, Toronto, ON M5S 3H4, Canada}
\affiliation{Department of Statistical Sciences, University of Toronto, 9th Floor, Ontario Power Building, 700 University Ave, Toronto, ON M5G 1Z5, Canada}
\affiliation{Data Sciences Institute, University of Toronto, 17th Floor, Ontario Power Building, 700 University Ave, Toronto, ON M5G 1Z5, Canada}

\author[0000-0003-2573-9832]{Joshua S. Speagle (\begin{CJK*}{UTF8}{gbsn}沈佳士\ignorespacesafterend\end{CJK*})}
\affiliation{Department of Statistical Sciences, University of Toronto, 9th Floor, Ontario Power Building, 700 University Ave, Toronto, ON M5G 1Z5, Canada}
\affiliation{David A. Dunlap Department of Astronomy \& Astrophysics, University of Toronto, 50 St George Street, Toronto, ON M5S 3H4, Canada}
\affiliation{Dunlap Institute for Astronomy \& Astrophysics, University of Toronto, 50 St George Street, Toronto, ON M5S 3H4, Canada}
\affiliation{Data Sciences Institute, University of Toronto, 17th Floor, Ontario Power Building, 700 University Ave, Toronto, ON M5G 1Z5, Canada}

\author[0009-0000-7936-4859]{Shu Yan Wang}
\affiliation{Department of Statistical Sciences, University of Toronto, 9th Floor, Ontario Power Building, 700 University Ave, Toronto, ON M5G 1Z5, Canada}

\begin{abstract}

The scaling relation between the size of a galaxy's globular cluster (GC) population ($N_{GC}$) and the galaxy's stellar mass ($M_*$) is usually described with a continuous, linear model, but in reality it is a count relationship that should be modeled as such. For massive galaxies, a negative binomial (NB) model has been shown to describe the data well, but it is unclear how the scaling relation behaves at low galaxy masses where a substantial portion of galaxies have $N_{GC}=0$. In this work, we test the utility of Poisson and NB models for describing the low-mass end of the $N_{GC}-M_*$ scaling relation. We introduce the use of \textit{zero-inflated} versions of these models, which allow for larger zero populations (e.g. galaxies without GCs) than would otherwise be predicted. We evaluate our models with a variety of predictive model comparison methods, including predictive intervals, leave-one-out cross-validation criterion, and posterior predictive comparisons. We find that the NB model is consistent with our data, but the naive Poisson is not. Moreover, we find that zero inflation of the models is not necessary to describe the population of low-mass galaxies that lack GCs, suggesting that a single formation and evolutionary process acts over all galaxy masses. Under the NB model, there does not appear to be anything unique about the lack of GCs in many low-mass galaxies; they are simply the low-mass extension of the larger $N_{GC}-M_*$ scaling relation.

\end{abstract}

\section{Introduction} \label{sec:intro}

Globular clusters (GCs) provide a wealth of information about their host galaxies. Strong correlations exist between the mass (or number) of GCs in a galaxy and other properties such as galaxy halo mass, velocity dispersion, and black hole mass \citep[e.g.,][]{Burkert2010, Harris2011, Harris2013, Harris2014, Forbes2018}. 

Beyond the Local Group, GC counts are often reported instead of more descriptive -- but harder to measure -- masses. Therefore, we have an acute interest in understanding the scaling relation between the \textit{number} of GCs belonging to a galaxy ($N_{GC}$) and the galaxy's other properties such as mass. These relationships are often characterized using linear regression models meant for continuous response variables \citep[e.g.][]{Harris2013, Burkert2020}. Count data are discrete, however, and a class of statistical models especially designed for these type of data exist. These models are a type of generalized linear model (GLM) that transform a continuous, linear response into a discrete one \citep[see][for a description of GLMs]{McCullaugh1983}. Count models predict integer responses that match the data instead of un-physical responses such as fractions of counts. In particular, they perform far better than continuous models in the estimation of mean and variance in the low-count regime. Regression models for counts remain generally underused in astronomy, though, despite the prevalence of count data in a variety of fields (X-ray photons, neutrinos, planet counts, etc.). 

\cite{deSouza2015} introduced the use of count models for GC populations and found that, across a wide galaxy mass range, $N_{GC}$ does not appear to be Poisson distributed against a number of galactic predictor variables, but a negative binomial (NB) model fits the data and its dispersion well. Follow-up work on this topic has not yet been done, and the prevailing count distribution used for GC populations remains a Poisson.

Additionally, the mass scaling relation $M_*-M_{GC}$, which is well studied, can inform our choices of count models to explore. The $M_*-M_{GC}$ relation is linear for Milky Way-sized and larger galaxies, but becomes uncertain with a large dispersion for dwarfs \citep{Harris2013, Bastian2020, Chen2023}. Scaling relations introduced by \cite{Eadie2022, Berek2023} that include a zero-generating process in the form of a hurdle model were found to be good fits for the low-mass end of this scaling relation. Hurdle models are a form of GLM that combine a continuous response (for example, a linear regression) with a zero-generating process (i.e. a logistic regression). This allows for the modeling of a continuous population that includes zero values without needing to manually remove zeros and study them separately using occupation fractions. Therefore, it is worth investigating whether the same holds true for the low-number end of $N_{GC}$. This might point to the need for an associated physical process that creates large numbers of low-mass dwarfs without GC populations. 

In this vein, we introduce the use of zero-inflated count models for GC count data for low-mass galaxies. Zero-inflated count models are similar to hurdle models in that they combine a count model with a separate zero-generating process. However, unlike hurdle models, the zeros in zero-inflated models can arise either from the zero-generating process or from the standard count model itself. The zero-generating process acts just to \textit{increase} the total number of expected zeros, or galaxies without GCs. We investigate whether zero-inflation is necessary to model GC populations in low-mass galaxies and what this can tell us about the underlying processes of GC creation and destruction. We do this entirely in a Bayesian inference framework. 

This paper is structured as follows: in Section~\ref{sec:data}, we describe our data. In Sections~\ref{sec:models} and \ref{sec:zeromodels} we define our chosen count models (Poisson and negative binomial) and their zero-inflated counterparts, respectively. Section 
\ref{sec:methods} introduces our inference methods. Section~\ref{sec:comp_testing} explains our model evaluation procedures, and Section~\ref{sec:results} presents our results. Section~\ref{sec:disc_conc} ends with a summary and concluding remarks. Throughout, we use $\ln$ to denote the natural logarithm $\log_e$, and $\log$ to denote $\log_{10}$. 

\vspace{-\baselineskip}
\section{Data} \label{sec:data}
We use a compilation of three data sources with a focus on dwarf galaxies, originally compiled in \cite{Eadie2022}. The sample consists of Local Group galaxies, nearby dwarfs, and Virgo cluster galaxies. A summary of these data is presented here, but for further details, see \cite{Eadie2022}.

The Local Group sample is an amalgamation of galaxy member lists from \cite{McConnachie2012, Lim2015, Simon2019, DrlicaWagner2020, Forbes2020}. The globular cluster information is taken from \cite{Harris2013, Lim2015, Forbes2018, Forbes2020}. In total, the Local Group sample contains 100 galaxies, of which 20 have GC populations. This sample is the most complete in the universe due to our ability to image faint GC systems in our own Local Group.

The nearby dwarf galaxies, along with GC system masses, come from \cite{Georgiev2009, Georgiev2010}. This sample consists of 39 galaxies, of which 31 have GCs. It focuses on isolated dwarfs outside of the Local Group. 

The Virgo cluster survey is taken from \cite{Peng2008, Jordan2009}. This sample consists of 93 galaxies with stellar masses $M_*<10^{11}M_\odot$, all of which have GC systems. We use this sample to anchor the upper-mass end of our model, since the previous two samples contain far more low-mass galaxies than Milky Way-sized ones. We only include galaxies up to about the mass of the Milky Way, though, since we are focused on the low-mass end of the GC scaling relation.

\begin{figure*} 
    \centering
    \includegraphics[width=0.70\textwidth]{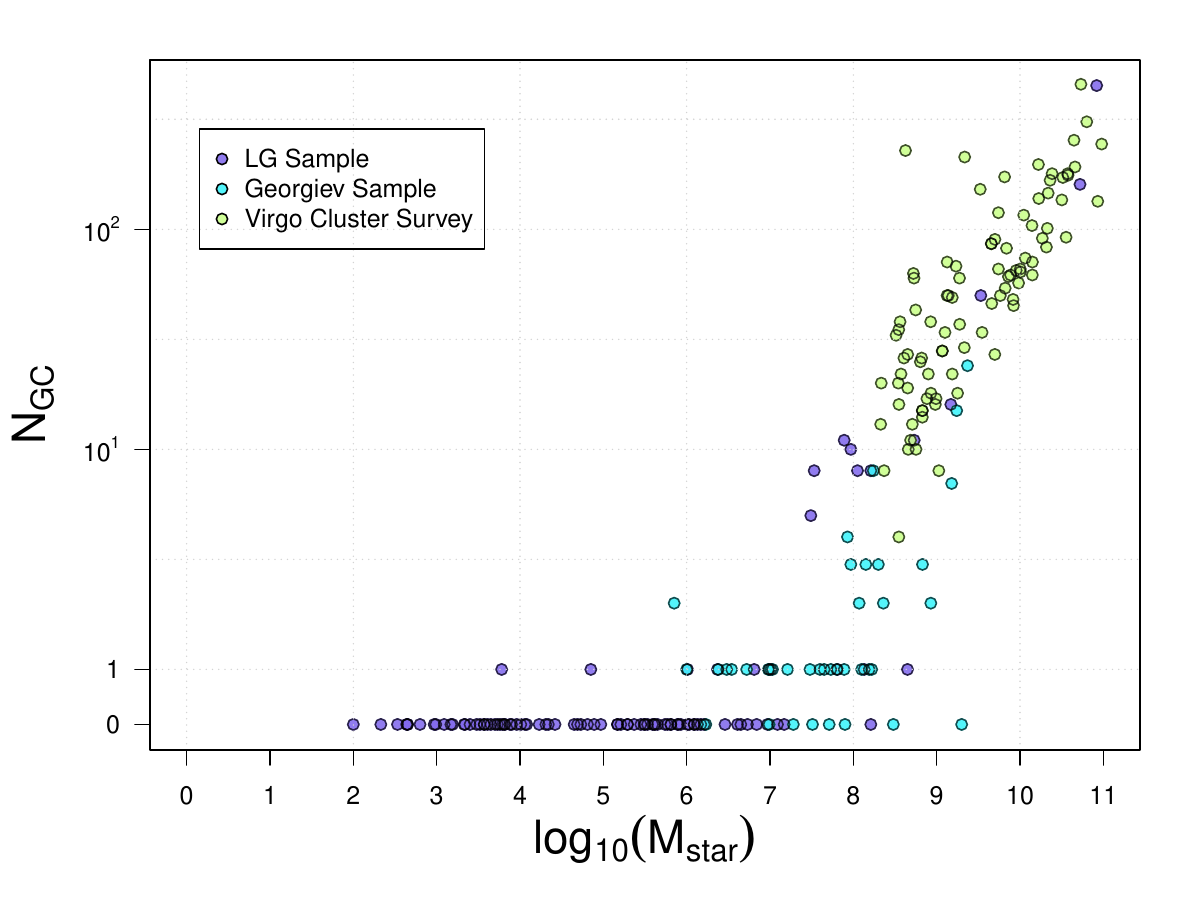}
    \caption{The number of GCs per galaxy, as a function of galaxy stellar mass, for our data. Points are colored according to the data source (i.e., Local Group sample, Georgiev sample, or Virgo Cluster Survey sample). The data consist of a large number of galaxies that do not have any globular clusters, as well as a substantial number of higher mass galaxies that have non-zero GC systems.}
    \label{fig:data}
\end{figure*}

Our data sample is plotted in Figure \ref{fig:data}. The galaxies span the stellar mass range $10^3<M_\odot<10^{11}$, from ultra-faint dwarfs to Milky Way-sized galaxies. About a third of the galaxies do not have a GC system, while the rest do. However, there is a large galaxy mass range for which these two populations overlap. 

\section{Count Models} \label{sec:models}

Count models are a class of generalized linear models (GLMs) that describe integer count data \cite[see][for a discussion of GLMs]{Nelder1972, McCullaugh1983, Gelman2013}. GLMs are linear in their parameters, but are mapped from this linear space to a different set via a link function $g(\mu)$ that transforms the mean response $\mu$.

GLMs have been introduced in only a few astronomy contexts (in addition to those mentioned in Section~\ref{sec:intro}), such as: modeling star formation and metal enrichment, and nuclear star cluster fractions, using binomial regression \citep{deSouza2015b, Zanatta2021}; photometric redshift estimation using gamma regression \citep{Elliott2015}; modeling ionizing radiation fractions using hurdle models with binomial and beta regression \citep{Hattab2019}; estimates of the dark matter field using negative binomial regression \citep{Ata2015}; and modeling galaxy cluster richness using Poisson regression \citep{Andreon2010}.

Here, we introduce two common count models: a Poisson model and a negative binomial model. 

\subsection{The Poisson model} \label{subsec:poisson}

The arguably simplest model for count data is the Poisson. The Poisson distribution describes the number of events that take place during some time period or within some spatial area, assuming independence between events and a constant mean rate of occurrence. It has only one parameter, $\lambda$, which describes both the mean and variance of the distribution. This implies that the mean is equal to the variance. In the context of galaxies and their GC populations, the number of GCs around a particular galaxy can be regarded as a realization of the random variable $Y$, which represents the number of events (GCs), given some predictor variable (in our case, log of galaxy stellar mass).

The probability mass function (PMF) $f(y)$ of the Poisson distribution gives the probability of count $y$, and is given by 
\begin{equation}
\label{eq:Poisson}
     f_\mathrm{P}(y) = \frac{\lambda^{y}}{y!}\exp{(-\lambda)}.
\end{equation}
where $y=0,1,2,\ldots$. Here, we have introduced the subscript $P$ to $f(y)$ denote the Poisson PMF.

To perform Poisson regression, the vector of means $\boldsymbol{\lambda}$ is related to a matrix of predictor variable(s) $\boldsymbol{X}$, called the design matrix\footnote{This term comes from applied statistics. Each row in the design matrix is an observation. The first column of the matrix is a series of 1's (or 0's), indicating the inclusion (or not) of the intercept term in the model. The subsequent columns are the covariates/predictors/features.}, through a link function $g()$:
\begin{equation} \label{equ:betas}
    g(\boldsymbol{\lambda}) = \boldsymbol{X} \boldsymbol{\beta}.
\end{equation}
The vector $\boldsymbol{\lambda}$ is thus a linear combination of the predictor $\boldsymbol{X}$ and a vector of coefficients $\boldsymbol{\beta}$, which is then transformed through the link function. In our application, the design matrix $\boldsymbol{X}$ is two columns wide: a column of 1s (for the intercept term) and a column for the covariate of galaxy stellar mass $M_*$. Thus, we fit a first-order linear combination: $g(\lambda_i) = \beta_{0} + \beta_{1} M_{*,i}$. The link function used for Poisson regression is the log link, such that $g(\boldsymbol{\lambda})=\ln{\boldsymbol{\lambda}}$.

The likelihood for Poisson regression is based on the Poisson PMF. With $\boldsymbol{\lambda}=\exp(\boldsymbol{X\beta})$ and with some algebraic simplifications, this leads to the log-likelihood:

\begin{equation}
    \ell_\mathrm{P}(\boldsymbol{y;\beta,X})=\sum_{i=1}^n\Big[(X_i\boldsymbol{\beta})^{y_i}-\exp(X_i\boldsymbol{\beta})-\ln \big(y_i!\big)\Big].
\end{equation}

In a standard Poisson distribution, the mean is equal to the variance (both are $\boldsymbol{\lambda}$) --- this is referred to as \textit{equidispersion}. However, there are cases when the variance is higher or lower than the mean, which is called \textit{overdispersion} and \textit{underdispersion}, respectively. The overdispersion case is much more common than the underdispersion case. There exists a class of count models, called overdispersed Poisson models, that explicitly handle data that is Poisson overdispersed. 

\subsection{The negative binomial model} \label{subsec:nb}
The negative binomial (NB) distribution is one of the most popular count models for overdispersed Poisson data. It is a mixture of Poisson and gamma distributions where the Poisson has a gamma-distributed mean. It therefore behaves similarly to a Poisson distribution, but also contains a second parameter to model variance.

There are multiple parameterazations of the NB model. We use the one described in \cite{Gelman2013}. This parameterization has parameters $\lambda$ and $\phi$, which relate to the mean and variance as: 

\begin{align}
    \mathrm{E}[Y] = \lambda \nonumber \\ 
    \mathrm{Var}[Y] = \lambda + \frac{\lambda^2}{\phi}.
\end{align}
In this parameterization, $\phi$ can be thought of as an ``overdispersion'' parameter that is inversely related to the variance. As $\phi$ decreases, the amount of extra dispersion compared to the Poisson dispersion $\lambda$ increases, and is scaled by $\lambda^2$. Often, $\phi$ is taken to be a constant that is not dependent on the predictor variables, but it can also be a linear function of the design matrix $\boldsymbol{X}$ with the form $g(\boldsymbol{\phi})=\boldsymbol{X\gamma}$.

The PMF of the NB model is 
\begin{equation}
\label{eq:NB}
    f_{\mathrm{NB}}(y) = \bigg(\frac{(y+\phi-1)!}{(\phi-1)!y!} \bigg) \bigg(\frac{\lambda}{\lambda+\phi}\bigg)^y \bigg(\frac{\phi}{\lambda+\phi}\bigg)^\phi,
\end{equation}
where $y=0,1,2,\ldots$.

Similarly to Poisson regression, NB regression also uses a log link function for $\boldsymbol{\lambda}$. This leads to the log-likelihood function:
\begin{multline} \label{eq:nblike}
    \ell_{\mathrm{NB}}(\boldsymbol{y;\beta},\phi,\boldsymbol{X})=\sum_{i=1}^n\bigg[\ln \big[(y_i+\phi-1)!\big] -\ln\big[(\phi-1)!y_i!\big] \\ 
    +y_iX_i\boldsymbol{\beta}+\phi\ln(\phi)
    -(y_i+\phi)\ln\big[\exp(X_i\boldsymbol{\beta})+\phi\big]\bigg],
\end{multline}
where we leave $\phi$ as a parameter that does not depend on the design matrix $\boldsymbol{X}$. However, if it does depend on $\boldsymbol{X}$, it is also transformed with a log link function $\ln(\boldsymbol{\phi})=\boldsymbol{X\gamma}$ in the likelihood. We explore models both where $\phi$ is a constant and where it is a first order function of the predictors $\ln(\phi_i)=\gamma_0+\gamma_1X_i$.

\section{Zero-Inflated Count Models} \label{sec:zeromodels}

When a data set has an excess of zero values compared to that predicted by a Poisson, NB, or other model, the data is considered \textit{zero-inflated}. Zero-inflated models describe scenarios where some of the zeros in a data set come from a process that also generates non-zeros (for example, a Poisson distribution), while others come from a separate process that only generates zeros.

A wide variety of models have zero-inflated counterparts. Here, we discuss the zero-inflated versions of the Poisson and NB models that were described in the previous section.

\subsection{The zero-inflated Poisson model} \label{subsec:zip}

The zero-inflated Poisson (ZIP) distribution was first described by \citet{lambert1992} and describes data for which there is an excess of zeros compared to a normal Poisson distribution. A ZIP model adds a Poisson PMF to a purely zero-generating process. This gives a mixture PMF of 
\begin{equation}
\label{eq:ZIP}
f_{\mathrm{ZIP}}(y)=
    \begin{cases}
        \pi_0 + (1-\pi_0)f_\mathrm{P}(y=0|\lambda) & \text{if } y=0 \\
        (1-\pi_0)f_\mathrm{P}(y|\lambda) & \text{if } y>0,
    \end{cases}
\end{equation}
where $f_\mathrm{P}(y)$ is the PMF of the Poisson distribution (Eq.~\ref{eq:Poisson}) and $\pi_0$ is the fraction of excess zeros. Under this model, some of the zeros are from the Poisson distribution, while others are from the zero-generating process.

ZIP regression is similar to Poisson regression, also using the log link function for the mean parameter $\boldsymbol{\lambda}$. The zero-inflation parameter $\boldsymbol{\pi_0}$ can be decomposed as a linear function of the predictor $\boldsymbol{\pi_0}=\boldsymbol{X\eta}$, allowing the probability of excess zeros to depend on the predictor variable(s). The link function for the zero-inflated parameter is a logit, such that $\ln\Big(\frac{\boldsymbol{\pi_0}}{1-\boldsymbol{\pi_0}}\Big)=\boldsymbol{X\eta}$ \citep[e.g.,][]{Yang2009, Campbell2021}. The (simplified) log-likelihood is therefore:
\begin{multline}
\ell_{\mathrm{ZIP}}(\boldsymbol{y;\beta,X,\eta})=
    \begin{cases}
        \sum\limits_{i=1}^n\bigg[\ln\Big[\Big(\frac{\exp(X_i\boldsymbol{\eta})}{1+\exp(X_i\boldsymbol{\eta})}\Big) \\
        \big(1-\exp(-\exp(X_i\boldsymbol{\beta}))\big)\\
        +\exp\big(-\exp(X_i\boldsymbol{\beta})\big)\Big]\bigg] & \text{if } y=0 \\ 
        \\
        \sum\limits_{i=1}^n\bigg[\ln\Big[\frac{1}{1+\exp(X_i\boldsymbol{\eta})}\Big]\bigg]\\
        +\ell_{\mathrm{P}}(y_i;\boldsymbol{\beta,}X_i) & \text{if } y>0.
    \end{cases}
\end{multline} 

\vspace{\baselineskip}
\subsection{The zero-inflated negative binomial model} \label{subsec:zinb}

The zero-inflated negative binomial (ZINB) distribution is an extension of the NB distribution in the same way that the ZIP distribution is an extension of the Poisson distribution. The ZINB distribution is a mixture of an NB with a zero-generating process, such that some zeros in the data come from the standard NB and others come from the zero process. The PMF is therefore:
    \begin{equation}
    \label{eq:ZINB}
f_{\mathrm{ZINB}}(y)=
    \begin{cases}
        \pi_0 + (1-\pi_0) f_{\mathrm{NB}}(y=0|\lambda,\phi) & \text{if } y=0 \\
         (1-\pi_0) f_{\mathrm{NB}}(y|\lambda,\phi) & \text{if } y>0,
    \end{cases}
\end{equation}
where $\pi_0$ is again the probability of excess zeros and $f_{\mathrm{NB}}(y)$ is the PMF of the negative binomial distribution (Eq.~\ref{eq:NB}). In regression, the parameters are transformed by a log link for $\boldsymbol{\lambda}$ (and $\boldsymbol{\phi}$ if it is a function of $\boldsymbol{X}$), and a logit link for $\boldsymbol{\pi_0}$. The log-likelihood function for ZINB regression is then:
\begin{multline} \label{eq:zinblike}
\ell_{\mathrm{ZINB}}(\boldsymbol{y;\beta,\eta,}\phi,\boldsymbol{X})=
    \begin{cases}
        \sum\limits_{i=1}^n\bigg[\ln\Big[\frac{\exp(X_i\boldsymbol{\eta})}{1+\exp(X_i\boldsymbol{\eta})} \\
        +\Big(\frac{1}{1+\exp(X_i\boldsymbol{\eta})}\Big)\\
        \Big(\frac{\phi}{\exp(X_i\boldsymbol{\beta})+\phi}\Big)^{\phi}\Big]\bigg] & \text{if } y=0 \\
        \\
        \sum\limits_{i=1}^n\bigg[\ln\Big[\frac{1}{1+\exp(X_i\boldsymbol{\eta})}\Big]\bigg]\\
        +\ell_{\mathrm{NB}}(\boldsymbol{y|\beta},\phi,\boldsymbol{X}) & \text{if } y>0.
    \end{cases}
\end{multline}
Again, we leave $\phi$ as a parameter that does not depend on the predictors in Eq. \ref{eq:zinblike}. In a version where it is a function of $\boldsymbol{X}$, it must be transformed with the log link function $\ln(\boldsymbol{\phi})=\boldsymbol{X\gamma}$ in the likelihood.

\section{Methods} \label{sec:methods}

We have chosen six different count models to test on our data set: 
\begin{enumerate}
    \item a Poisson
    \item a zero-inflated Poisson (ZIP)
    \item a negative binomial (NB)
    \item an NB with a variable dispersion parameter (dependent on $M_*$)
    \item a zero-inflated NB (ZINB)
    \item a ZINB with a variable dispersion parameter (dependent on $M_*$).
\end{enumerate}
Models with a non-constant dispersion parameter allow for different amounts of dispersion about the mean in different ranges of galaxy mass, which could correspond to the impacts of different physical processes acting on GCs in massive vs low-mass galaxies.

\subsection{Inference}

We use Bayesian inference for our modeling. Bayesian inference is based on Bayes' theorem, which states that
\begin{equation}
    P(\theta|y)=\frac{P(y|\theta) P(\theta)}{P(y)},
\end{equation}
where $\theta$ are the model parameters and $y$ are the data. 
$P(\theta|y)$ is the posterior, or the probability of a certain $\theta$ given $y$. $P(y|\theta)$ is the likelihood, $P(\theta)$ is the prior, and $P(y)$ is the evidence, which is a normalization term.

We do not have much physical intuition for the priors in these models. Therefore, we follow the example of \citet{deSouza2015, Eadie2022} and use non-informative, broad priors. We choose the default priors in \texttt{brms} for each of the models, and data is centered before being fit. 

All models are run using the R statistical software environment \citep{R} and the package \texttt{brms} \citep{brms}. The \texttt{brms} package compiles and runs Bayesian models in Stan \citep{stan}, which is a Hamiltonian Monte Carlo (HMC) sampler \citep{hmc, nuts}. HMC algorithms are known to be more efficient than typical random walk samplers.

\subsection{Measurement uncertainties}

Ideally, one would include uncertainties for both galaxy stellar masses and GC counts in the analysis. Unfortunately, we do not have uncertainties for all of our data. While uncertainties on stellar masses are reported for the Local Group and Virgo samples, they are not reported for the Georgiev sample. Furthermore, none of the samples report uncertainties in GC counts, even though we expect uncertainties to exist. For example, we could be missing GCs that are located behind galaxy disks or that are fainter than detection limits, and counts could contain contamination from background galaxies in regimes where GCs appear as point sources (i.e. the Virgo sample).

While we could make some assumptions about the missing uncertainties and impute them, it is also not trivial to account for uncertainties in count models. Incorporating uncertainties in predictor variable(s) requires an errors-in-variables model, which adds $N$ parameters to the fit \citep[see][for an example of errors-in-variables models in astronomy]{Berek2023}. Incorporating uncertainties in a count response variable relies on  an unknown PMF that includes selection effects and detection noise to link observed counts to true counts. This kind of uncertainty analysis is beyond the scope of this work and may require novel statistical methods. Thus, we leave this to future work and instead rely on the fact that our sample consists of nearby galaxies and thus we expect their GC data to be relatively complete. High levels of completeness are also suggested by two of our three data sources which have conducted completeness analyses on their samples \citep[see][for details of their completeness analyses]{Georgiev2009, Jordan2009}.

\section{Model Evaluation and Comparison} \label{sec:comp_testing}

There exist many methods to perform model comparison and evaluation. In astronomy, information criterion like the Akaike information criterion (AIC) and Deviance information criterion (DIC) are popular \citep{Akaike1973, Spiegelhalter2002}. These methods give models a score based on the maximum likelihood estimate of the parameters (in AIC) for frequentist models, or based on the posterior mean point estimate of the parameters (in DIC) for Bayesian models. A fully Bayesian approach is the Watanabe-Akaike information criterion (WAIC), which computes the log average likelihood for each data point across the entire posterior distribution \citep{Watanabe2010}. However, all of these methods rely on the maximum likelihood estimate or posterior of the model fit to the \textit{original} data, instead of evaluating how well the model can \textit{predict} additional data. The ability to accurately predict new data is a more philosophically Bayesian method of evaluating a model, and is practically useful in many astronomy cases, such as simulation building. 

In this section, we describe multiple methods of predictive model comparison, which we will then apply on our models in Section \ref{sec:results}. We use three different predictive methods that evaluate and compare models in different ways: (1) predictive intervals, which compare the intervals of real data to data simulated from our model; (2) leave-one-out cross validation, which compares the probability of a new, unseen data point across different models, and (3) a posterior predictive comparison, which evaluates the probability of the model itself.

\subsection{Predictive Intervals}
Predictive intervals show the range within which future data are expected to fall. This is different from a credible interval, which shows the range of uncertainty in the model parameters. Therefore, unlike credible intervals, predictive intervals provide an intuitive way of comparing not only the mean, or expectation value, of the model, but also the dispersion about the mean. For example, if the model is a good description of the data, then we would expect roughly 75\% of the data to fall within the 75\% prediction interval. 

\subsection{Leave-one-out cross validation (LOO-CV)} \label{subsec:loocv}

LOO-CV \citep{Vehtari2017} estimates the predictive abilities of a model on new data by removing one data point ($y_i$) and refitting the model to the $n-1$ remaining points ($y_{-i}$). It relies on the assumption that the fit to $y_{-i}$ should be similar to the original fit to $y$, which allows for a predictive assessment without requiring additional data. LOO-CV calculates the expected log predictive density, or $P_{y_i}$, which is the probability of predicting the removed point $y_i$ based on the fit to $y_{-i}$. This is analogous to the log likelihood, and is given by:
\begin{equation}
    P_{y_i} = \sum_{i=1}^{n} \log{p(y_i|y_{-i})}
\end{equation}
for each removed data point $y_i$ and remaining data $y_{-i}$ on which the model is fit, where
\begin{equation}
    p(y_i|y_{-i}) = \int p(y_i|\theta, y_{-i})p(\theta|y_{-i})d\theta
\end{equation}
is the leave-one-out posterior predictive density for the $y_{-i}$ data. To calculate a total LOO-CV value, the $P_{y_i}$ for each removed point $y_i$ are combined into a composite score.

$P_{y_i}$ is a measure of how well the model can predict new data ($y_i$) that is not included in the model fit. A higher value indicates that more probability is concentrated at the unknown points $y_i$, making them more likely. However, these values are typically multiplied by $-2$ to follow the convention of other commonly used model comparison methods such as the AIC, DIC, and WAIC. Therefore, lower LOO-CV scores indicate better models. 

\subsection{Posterior Predictive Comparison} \label{subsec:elpd}

\begin{figure*} 
    \centering
    \includegraphics[width=0.99\textwidth]{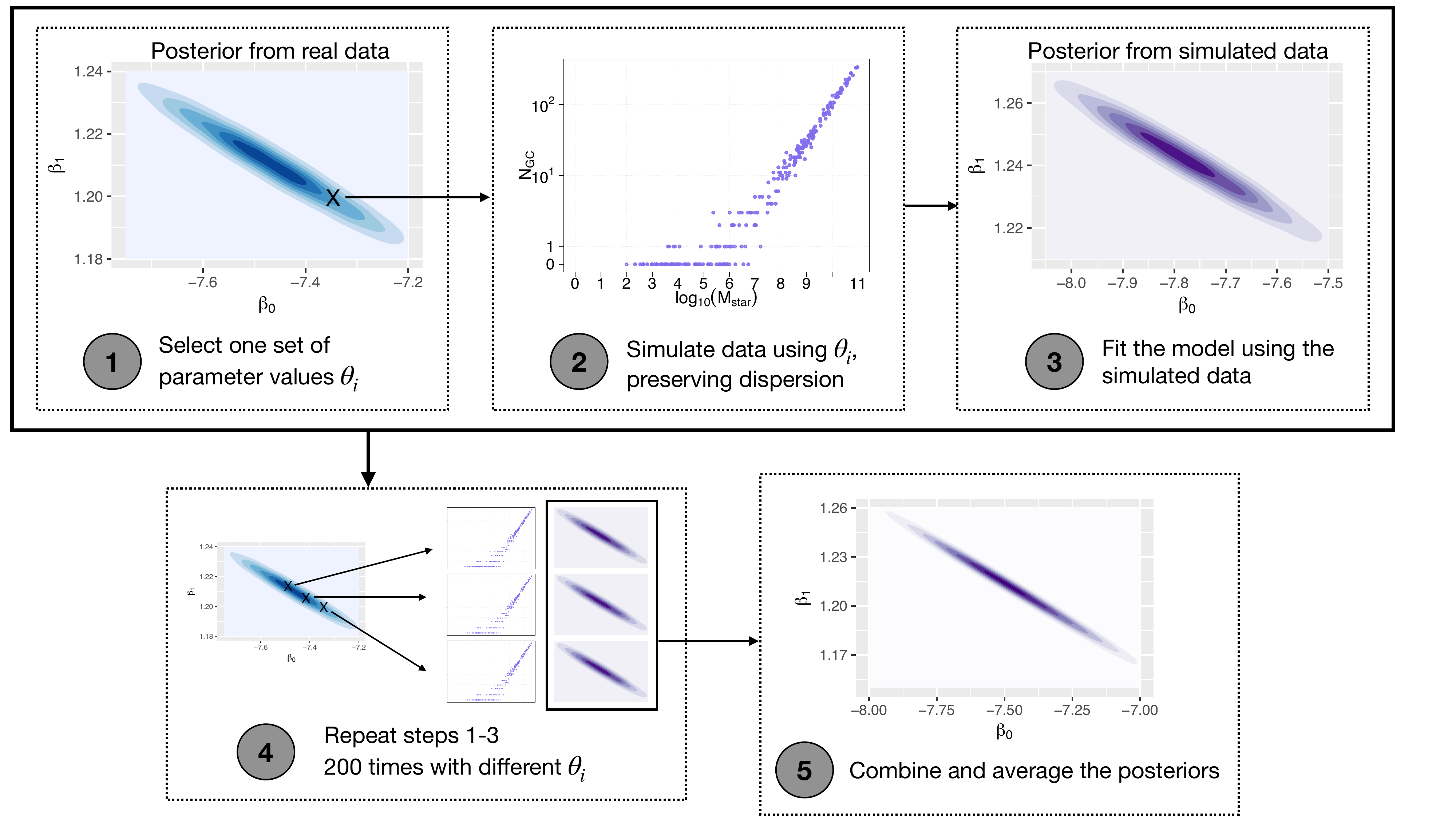}
    \caption{The process of computing the predicted range of posteriors for each model. A simple, two parameter model is illustrated here for visualization purposes. In step 1, one chain of the MCMC used to fit the real data to the model, with parameters $\theta_i$, is selected. Next, we use the $\theta_i$ to simulate data with the same galaxy masses and dispersion as the original data. In step 3, we re-run the MCMC with this simulated data. We repeat this process 200 times with different chains $\theta_i$ and average the posteriors from the simulations. This gives the range of posteriors that are plausible for data that does fit the model, which can be compared to the posteriors from the fit with real data.}
    \label{fig:pred_schematic}
\end{figure*}

While LOO-CV is a useful model comparison method, it is only that. It can tell us which model is \textit{best}, but not if that model is \textit{good}. To evaluate whether any one model is consistent with the data, we use a posterior predictive check, which evaluates the likelihood of the posterior density instead of the comparing the likelihood of future data. To compute this posterior predictive, we do the following:
\begin{enumerate}
    \item Randomly select one set of parameter values $\theta_i$ from the estimate of the posterior (i.e. the Markov chains).
    \label{list:1}
    \item Given $\theta_i$, simulate $N_{GCS}$ for each $M_*$ in our data set. Add random noise $\epsilon \sim N(0, \mathrm{Var}(y))$ to $N_{GCS}$ to mimic the dispersion of the data.
    \label{list:2}
    \item Re-fit the model with these simulated data, obtaining a new estimate of the posterior distribution and the associated posterior probability density.
    \label{list:3}
    \item Repeat steps~\ref{list:1}-\ref{list:3} 200 times. Combine the posterior samples and re-normalize to create a ``global" posterior density.
    \item Compare the ``global" posterior density to the posterior density estimated from the observed data.
\end{enumerate}
These steps to compute the predictive posterior are illustrated in Figure \ref{fig:pred_schematic}.

This predictive check provides a direct comparison between the ``global" posterior density [i.e. the range of possible posteriors for data generated from a given model], and the posterior from our real data. If the posterior of our data falls within the range of expected posteriors, this indicates that our data is consistent with our model. 

\section{Results} \label{sec:results}

The estimated parameter values for each of the six models are listed in Table \ref{tab:params}. To calculate estimates of the expectation value (mean), dispersion, and zero-inflation, the parameter values need to be transformed as indicated in Table \ref{tab:eqs}, based on the link functions used for regression.

\vspace{-\baselineskip}
\vspace{-\baselineskip}
\vspace{-\baselineskip}
\vspace{-\baselineskip}
\begin{deluxetable*}{cccccccc}[t] \label{tab:params}
    \tablecaption{Posterior mean parameter values for each model.}
    \tablehead{
    \colhead{Model Version} & \colhead{$\beta_{0}$} &  \colhead{$\beta_{1}$} & \colhead{$\eta_{0}$} &  \colhead{$\eta_{1}$} & \colhead{$\phi$} &  \colhead{$\gamma_{0}$} & \colhead{$\gamma_{1}$}
    }
    \startdata 
         Poisson & $-7.46$ & $1.21$ & -- & -- & -- & -- & -- \\
          & $(-7.72,-7.21)$ & $(1.18,1.24)$ & -- & -- & -- & -- & -- \\
          \\
         ZIP & $-6.92$ & $1.16$ & $8.32$ & $-1.28$ & -- & -- & --\\
         & $(-7.20,-6.63)$ & $(1.13,1.19)$ & $(5.20,11.69)$ & $(-1.74,-0.87)$ & -- & -- & -- \\
         \\
         NB & $-10.35$ & $1.52$ & -- & -- & $1.29$ & -- & -- \\
         & $(-11.57,-9.16)$ & $(1.38,1.65)$ & -- & -- & $(1.00,1.64)$ & -- & -- \\
         \\
         NB & $-9.30$ & $1.40$ & -- & -- & -- & $-4.19$ & $0.49$ \\
         (non-constant disp.) & $(-10.61,-7.90)$ & $(1.25,1.54)$ & -- & -- & -- & $(-6.41,-1.71)$ & $(0.22,0.73)$ \\
          \\
         ZINB & $-10.03$ & $1.48$ & $1.51$ & $-0.70$ & $1.37$ & -- & -- \\
          & $(-11.40,-8.65)$ & $(1.33,1.63)$ & $(-10.18,9.10)$ & $(-1.76,0.49)$ & $(1.04,1.77)$ & -- & -- \\
          \\
         ZINB & $-9.11$ & $1.38$ & $1.11$ & $-0.69$ & -- & $-4.04$ & $0.48$ \\
         (non-constant disp.) & $(-10.58,-7.66)$ & $(1.23,1.54)$ & $(-9.45,8.78)$ & $(-1.78,0.50)$ & -- & $(-6.30,-1.66)$ & $(0.22,0.73)$\\
    \enddata
    \tablecomments{The values listed are the posterior means for each parameter of each model. The values in parentheses are the 95\% credible intervals. Linear combinations of the above parameters, transformed by their respective link functions, provide estimates for the parameters $\lambda$, $\phi$, and $\pi_0$. See Table \ref{tab:eqs} for the equations needed to transform these fitted parameters into count estimates.}
    \vspace{-6mm}
\end{deluxetable*}

\begin{deluxetable*}{cccc}[t] \label{tab:eqs}
    \tablecaption{How to use the values in Table \ref{tab:params}.}
    \tablehead{
    \colhead{} & \colhead{$\mathrm{E}[N_m]=\lambda$} &  \colhead{$\mathrm{Var}[N_m]=SD[N_m]^2=\lambda+\frac{\lambda^2}{\phi}$} & \colhead{$\pi_0$} \\
    \colhead{} & \colhead{mean} & \colhead{variance} & \colhead{fraction of zero-inflation}}
    \startdata 
         Poisson & $\exp{(\beta_0+\beta_1m)}$ & -- & -- \\ 
         Example ($\log m=10)$ & $ 104$ & -- & -- \\ \\ 
         ZIP & $ \exp{(\beta_0+\beta_1m)}$ & -- & $\frac{\exp(\eta_0+\eta_1m)}{1+\exp(\eta_0+\eta_1m)}$ \\
          Example ($\log m=10)$ & $ 108$ & -- & $0.0112$ \\ \\
         NB & $ \exp{(\beta_0+\beta_1m)}$ & $ \exp(\beta_0+\beta_1m)+\frac{(\exp(\beta_0+\beta_1m))^2}{\phi}$ & -- \\
          Example ($\log m=10)$ & $ 128$ & $ 113^2$ & -- \\ \\
         NB (non-constant disp.) & $ \exp{(\beta_0+\beta_1m)}$ & $ \exp{(\beta_0+\beta_1m)}+\frac{(\exp(\beta_0+\beta_1m))^2}{\exp{(\gamma_0+\gamma_1m)}}$ & -- \\ 
          Example ($\log m=10)$ & $ 110$ & $ 78^2$ & --\\ \\
         ZINB & $ \exp{(\beta_0+\beta_1m)}$ & $ \exp{(\beta_0+\beta_1m)}+\frac{(\exp(\beta_0+\beta_1m))^2}{\phi}$ & $\frac{\exp(\eta_0+\eta_1m)}{1+\exp(\eta_0+\eta_1m)}$ \\ 
          Example ($\log m=10)$ & $ 118$ & $ 101^2$ & $0.0041$\\ \\
         ZINB (non-constant disp.) & $ \exp{(\beta_0+\beta_1m)}$ & $ \exp{(\beta_0+\beta_1m)}+\frac{(\exp(\beta_0+\beta_1m))^2}{\exp{(\gamma_0+\gamma_1m)}}$ & $\frac{\exp(\eta_0+\eta_1m)}{1+\exp(\eta_0+\eta_1m)}$ \\ 
          Example ($\log m=10)$ & $ 118$ & $ 81^2$ & $0.0030$
    \enddata
    \tablecomments{How to calculate the mean, variance, and zero-inflation for each model from the posterior mean values in Table \ref{tab:params}. $m$ refers to the log of the stellar mass for which a parameter value is to be calculated (i.e. $m=\log M_*$), and $N_m$ refers to the number of GCs corresponding to the chosen $m$. The $\pi_0$'s for all models are poorly constrained (see the credible intervals in Table \ref{tab:params}) and so the values calculated for the example are not considered robust.}
    \vspace{-6mm}
\end{deluxetable*}

\subsection{Predictive intervals}

\begin{figure*} 
    \centering
    \includegraphics[width=0.99\textwidth]{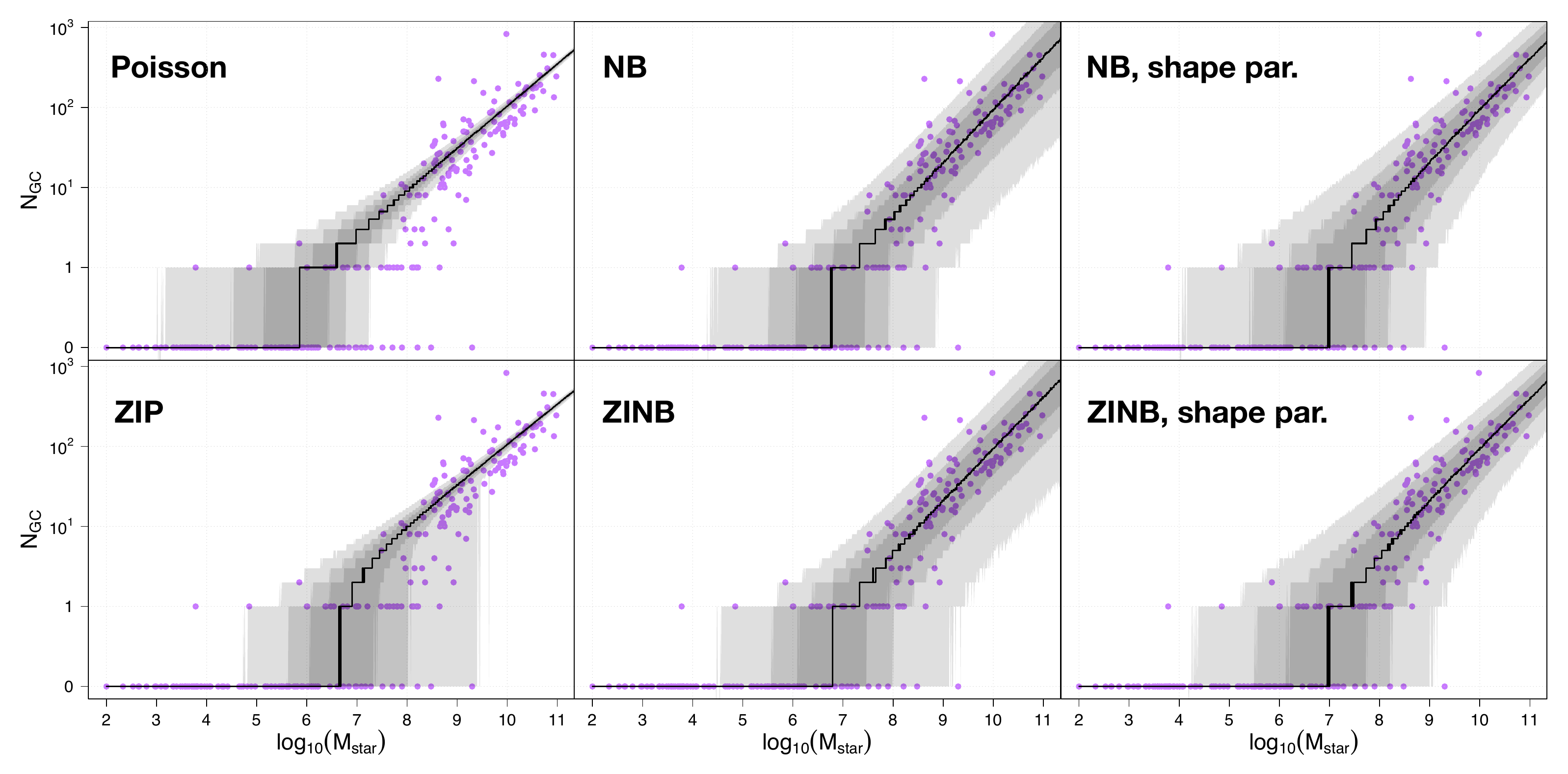}
    \caption{Predictive intervals for the six models. Shaded regions, from outer to inner, show the ranges in which 95\%, 75\%, and 50\% of new data would be expected to lie. These regions should also encompass 95\%, 75\%, and 50\% of the plotted galaxy data, if the given model is a good fit to the data.}
    \label{fig:pred_int}
\end{figure*}

Predictive intervals for each of the six models are shown in Figure \ref{fig:pred_int}. The prediction intervals are, from outer to inner, 95\%, 75\%, and 50\%. Black lines show the mean predictions, and purple points are our data. 

Both versions of Poisson models shown in Figure \ref{fig:pred_int} fail to capture the dispersion of the data, especially for larger galaxy masses. The ZIP model is better at predicting the mass range of galaxies without GCs, but both models do equally poorly at higher galaxy masses. All four versions of the NB model have much larger dispersions than the Poisson, especially at higher galaxy mass. The NB models with a dispersion parameter dependent on $M_*$ have a dispersion that decreases with increasing galaxy mass. However, the varying dispersion of these models do not significantly change the percentage of our data that fall within any of the intervals, compared to the non-variable dispersion case. The 95\% predictive intervals, for example, cover about 95\% of the data in all four of the NB models, meaning that they all seem capable of producing data similar to our galaxy sample. 

\vspace{-\baselineskip}
\subsection{LOO-CV}

To more quantitatively assess model fit, we compare our six models using LOO-CV, which is computationally simple in \texttt{brms} and Stan using the \texttt{loo} package \citep{loo}. As discussed in Section \ref{subsec:loocv}, lower values indicate a better model. 

The LOO information criterion values for each model are listed in Table \ref{tab:loocv}, along with their standard errors. The four versions of the NB model outperform the Poisson and ZIP. The differences between the various NB models, however, are indistinguishable given their standard errors. As discussed in \citet{Gelman2013}, metrics like LOO-CV are only one tool in the model comparison and evaluation toolbox. Best practice suggests using multiple methods of model comparison and evaluation, along with physical intuition about the data and model choices, before coming to a conclusion.

\begin{deluxetable*}{ccc} \label{tab:loocv}
    \tablecaption{LOO-CV values for all models.}
    \tablehead{
    \colhead{Model Version} & \colhead{LOO Information Criterion} &  \colhead{Standard Error}
    }
    \startdata 
         Poisson & 6046.4 & 2245.1 \\
         ZIP & 5862.7 & 2208.0 \\
         Negative Binomial & 1323.9 & 82.8 \\
         ZINB & 1327.8 & 83.5 \\
         Negative Binomial (non-constant dispersion) & 1318.1 & 83.4 \\
         ZINB (non-constant dispersion) & 1321.3 & 84.1 \\
    \enddata
    \tablecomments{Smaller LOO information criterion values indicate more likely models.}
    \vspace{-3mm}
\end{deluxetable*}

\subsection{Posterior Predictive Comparison}

\begin{figure*} 
    \centering
    \includegraphics[width=0.99\textwidth]{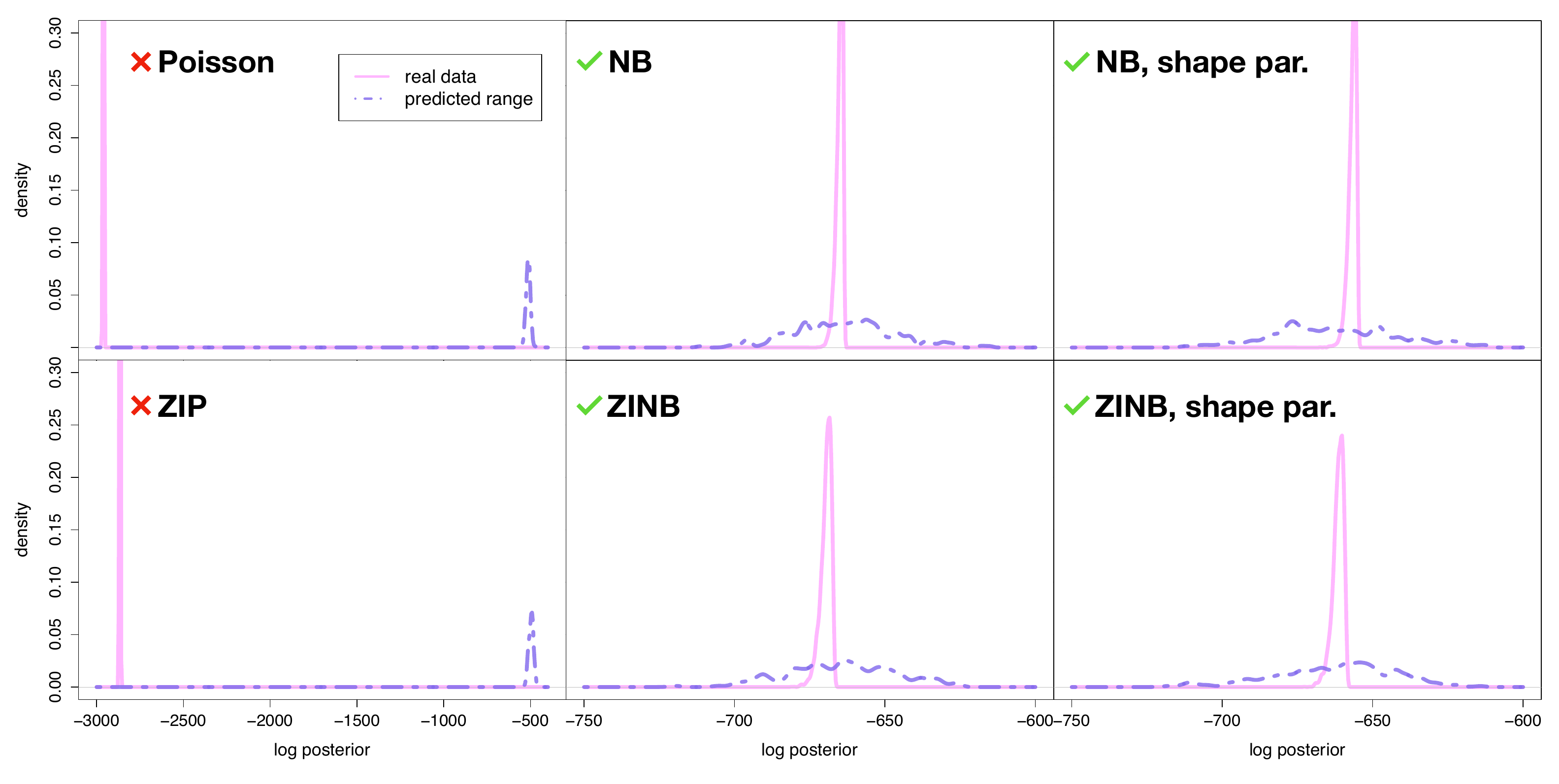}
    \caption{Posterior predictive comparisons for all six models. The solid pink posteriors are those from the real data fit to the model, and the dashed purple posteriors are the average from the posterior predictive check; i.e. the model run on data simulated from draws from the original posterior. Note: the y-axes are cut off at a density of 0.4 for visual clarity.}
    \label{fig:post_pred}
\end{figure*}

We follow the procedure described in Section \ref{subsec:elpd} to perform a posterior predictive comparison, and plot the results in Figure \ref{fig:post_pred}. The ``global" posterior densities from the simulated data tests (in purple) show the range of possible posteriors that would be expected from data that is consistent with the model. The posteriors from our data (in pink) are one realization, and thus expected to be much narrower than the simulated distribution. If the ``global" and sample posterior distributions do not overlap, it indicates an unlikely sample posterior, i.e., that the data are not well fit to that model.

For the two Poisson distributions (first column in Figure~\ref{fig:post_pred}), the posteriors from the real data are significantly different than the range of possible posteriors suggested by the simulated data. This suggests that our data is not Poisson distributed. The four NB posteriors of the real data, however, overlap with the distribution of simulated posteriors. This means that the data are consistent with the NB models. Similarly to the LOO-CV results, there are no distinguishing factors between the four NB models. 

\vspace{\baselineskip}
\vspace{\baselineskip}
\vspace{\baselineskip}
\section{Discussion and Conclusions} \label{sec:disc_conc}

In this paper, we explore a variety of count models for use in GC counts of low-mass galaxies. We emphasize statistically robust methods of model comparison that are based in predictive methods, or the model's ability to predict new data not used in model fitting. Our primary conclusions are as follows: 
\begin{enumerate}
    \item The GC populations of low-mass galaxies are not well described by a Poisson regression. 
    \item NB regressions are consistent with the GC data, and indicate greater variance (overdispersion) in GC counts as a function of galaxy mass than would be expected by a Poisson.
    \item Zero-inflation is not necessary for GC populations, indicating a single process of GC formation over all galaxy masses.
\end{enumerate}
These insights regarding our understanding of large star clusters in the smallest galaxies can provide important constraints in simulations and theoretical work on cluster and galaxy formation and evolutionary processes. They are described in more detail below, and we end with a look toward the future of this field.

\subsection{GC counts are not Poissonian}
The most simple count-generating process is a Poisson process, in which points are independently generated by sampling from a density distribution. Globular cluster counts have often been modeled with Poisson models before \citep[e.g.][]{Pfeffer2018, Huang2021, Eadie2022}. In their analysis, \cite{Eadie2022} presented evidence for the first time that the Poisson was a poor representation of the GC systems of low-mass galaxies. We come to a similar conclusion. The data do not match the Poisson prediction intervals, LOO-CV shows the Poisson models to be much worse than the negative binomials, and the predictive comparisons confirm that the Poisson models are not consistent with the data. This leads us to two possible interpretations: (1) if we believe that GCs should have originally formed around galaxies in numbers that followed a Poisson regression with respect to galaxy host mass, then our results indicate there are physical processes at play that cause deviations from the Poisson over time (i.e., cluster evolution), or, (2) perhaps GCs do not form around galaxies in numbers consistent with a Poisson regression to begin with. 

A Poisson distribution is generated when events are sampled from the underlying density \textit{independently}. However, this is clearly not the case with globular cluster formation. There is a finite amount of gas in a galaxy, and when some of it is turned into a large star cluster, this removes that gas from the remaining gas that would be available to create more star clusters. This effect is more prominent in smaller galaxies, since they have less gas to begin with, and the formation of each large star cluster requires a larger percentage of the total gas in the galaxy. This galaxy-scale feedback from the formation of GCs violates the independence assumption of a Poisson distribution, making it an un-physical choice.

\subsection{An NB model provides insight for GC populations}

Just as \cite{deSouza2015} found negative binomial models to be a better fit to GC count data for large galaxies, we find the same for low-mass galaxies. Our LOO-CV model comparison indicates that NB models are a far better choice than the Poisson for this data, and this is confirmed by our posterior predictive checks. Our data are consistent with multiple realizations of data drawn from the negative binomial model. 

We began this model testing without much physical intuition for our models or priors. Our model comparison and predictive checks have informed us that NB models - but not Poisson models - are consistent with our sample of GCs of low-mass galaxies, which indicates that GC counts are overdispersed from a Poisson. The overdispersion corresponds to high variation in GC population size per galaxy mass. 

The large dispersion in GC counts as a function of galaxy mass could be caused by other covariates, such as central black hole mass, velocity dispersion, halo mass, or star formation histories, that are absent in our model. Some of these covariates have been shown to correlate strongly with GC populations \citep[e.g.][]{Burkert2010, Harris2011, Harris2013, Harris2014, deSouza2015, Forbes2018}, although it is unclear whether their correlation to galaxy stellar masses or to GC populations is stronger. Some of the overdispersion could also be caused by stochastic processes or other confounding variables that have not yet been linked to GC populations.

Further factors, such as environment, could also impact the size of GC populations. This type of group-level effect could be incorporated into a model through the use of partial pooling or mixed models, which have the ability to fit for different values of parameters within different subsets of the data (i.e. isolated vs cluster galaxies). If group-level effects are important in this relationship, our ignoring of these effects would result in higher dispersion. Regardless of the cause of the overdispersion, however, we stress that the NB model is a good \textit{empirical} model for the $M_*-N_{GC}$ relation. We leave the exploration of other covariates and mixed models to future work.

\vspace{\baselineskip}
\subsection{Zero-inflation is not necessary for GC populations}

\cite{Eadie2022} and \cite{Berek2023} both showed that a significant portion of the lowest-mass galaxies do not have any GCs. Thus, we introduced zero-inflated versions of all of our models to allow for excess zeros. The posterior mean of the zero-inflation ($\pi_0$) parameter of the ZIP model approaches 1 in the lowest-mass regime and 0 in the higher mass regime ($\pi_0=0.87$ for $\log M_*=5$ and $0.01$ for $\log M_*=10$), which indicates significant zero-inflation for the smallest galaxies, in line with expectations from other studies \citep[e.g.][]{Chen2023, Berek2023}. However, the ZIP model's LOO-CV score was equally as bad as that of the Poisson model. Therefore, we conclude that the $\pi_0$ parameter is attempting to compensate for the poor fit of the Poisson, rather than indicating a genuine need for a model with excess zeros. The zero-inflated NB models both have very poorly constrained $\pi_0$ parameters, and model comparison tests conclude that adding zero-inflation does not significantly improve the model fits. We operate under the philosophy that the simplest model that can adequately explain the data is the best one, and so we do not favor the zero-inflated (or variable dispersion NB) models over the simplest NB model. 

Therefore, although there is an increasing proportion of galaxies that do not have GCs with decreasing mass, these galaxies seem to be the natural extension of one overall process of GC formation and evolution that operates over all galaxy masses. In other words, the same rules of GC formation and evolution seem to extend to all galaxies, regardless of their mass. This is an important result, not only for physical interpretation, but also for simulation studies; if one wishes to generate a realistic number of GCs around a simulated galaxy of a particular mass, then we recommend using our NB parameters to simulate that number of GCs, given the stellar mass of the host galaxy. 

\subsection{Looking to the future}
The negative binomial model is consistent with the GC populations of nearby dwarf galaxies, but current GC counts in these galaxies are an amalgamation of the number of GCs that were originally formed and the number that have been completely disrupted due to mass-loss processes. GC formation and evolution are governed by different physical processes happening over different timescales, and thus may not follow the same statistical models. If we could remove cluster evolution as a factor, perhaps by looking at nearby massive cluster formation \citep{BerekReinaCampos2023}, newly formed GCs at high redshift in simulations such as E-MOSAICS and EMP-Pathfinder \citep{Pfeffer2018, ReinaCampos2022}, or younger GCs at high redshift with new telescopes like JWST \citep{Mowla2022, Adamo2024}, we could learn about cluster formation and evolution, and the statistical models that describe them, separately. 

It remains to be seen what factors physically motivate an NB model. The open questions surrounding GC formation and evolution at high redshifts prevent us from choosing a model based purely off of physical intuition, and so we rely on the data to drive our model selection instead. The NB model, though, may yet inform our physical understanding of these processes. GC counts are often highly uncertain, especially at farther distances, due to background contamination and observational limits. Comparisons of NB models which are known to describe high-quality GC data to lower-quality data sets can provide information about clusters we are missing and further our understanding of large-scale GC properties and galaxy evolution.

\section*{Acknowledgements}

The authors would like to thank the anonymous referee, whose comments greatly improved this manuscript. SCB would like to thank Steffani Grondin and Ayush Pandhi for helpful discussions throughout the writing of this manuscript. SCB is supported by the Data Sciences Institute at the University of Toronto through grant number DSI-DSFY3R1P24. GME acknowledges funding from NSERC through Discovery Grant RGPIN-2020-04554. JSS acknowledges funding from NSERC through Discovery Grant RGPIN-2023-04849. SYW was supported by the Summer Undergraduate Research Award (SURA) from the Department of Statistical Sciences, University of Toronto.

\software{Stan Modeling Language \citep{stan}, R Statistical Software Environment \citep{baser}, and the following R packages: \texttt{rstan} \citep{rstan}, \texttt{loo}, \cite{loopackage} \texttt{plotrix} \citep{plotrix}, \texttt{ggplot2} \citep{ggplot}, \texttt{bayesplot} \citep{bayesplot}.}

\vspace{5mm}

\bibliography{paper}{}
\bibliographystyle{aasjournal}

\end{document}